# Inversions in astronomy and the SOLA method

*Frank P. Pijpers*
Uppsala Astronomical Observatory
box 515, S-751 20 Uppsala, Sweden
INTERnet : Frank.Pijpers@astro.uu.se

**Abstract**

A brief overview of applications of inversions within astronomy is presented here and also an inventory of the techniques commonly in use. Most of this paper is concerned with a presentation of a recent modification of the well-known Backus & Gilbert method (1967, 1968, 1970).

In general inversions in astronomy arise when observational (experimental) data are a convolution of some quantity of astrophysical interest and a known or measured effect. The latter can be a known property of the instrument used for the observation, an effect of projection on the sky or, as in helioseismology, a convolution along the ray path of a seismic wave in the Sun. Since the measured data is sampled discretely and suffers from measurement errors of various kinds, it is rare that an exact analytical inversion can be carried out. Furthermore what distinguishes astronomy from most other experimental physical sciences is that both the sampling and the data errors are difficult or impossible to control. A number of numerical inversion techniques are currently in use that try to deal with these difficulties in various ways. A particularly useful reference that describes the basics of many inversion and deconvolution techniques are two chapters in the book by Press et al. (1992). Almost all the techniques described there have been applied in astronomy in one form or another.

The second part of this paper is focussed on a small selection of astronomical inversion problems, where the method of Subtractive Optimally Localized Averages (SOLA) has been used. The SOLA method is an adaptation of the Backus & Gilbert method (1967, 1968, 1970). It was originally developed for application in helioseismology (Pijpers & Thompson, 1992, 1993b, 1994) where the Backus & Gilbert method is computationally too slow. Apart from achieving a considerable speedup in the SOLA formulation, the strength of the method lies in that it provides a good a priori estimate of the error due to data error propagation and similarly a good a priori estimate of the achievable resolution. The latter property in particular turns out to be of importance in the problem of reverberation mapping of active galactic nuclei (Pijpers & Wanders, 1994). The freedom to choose a desired resolution within SOLA is also particularly useful if the 'known' function under the integral sign is a measured quantity with associated measurement noise, as in reverberation mapping, because it allows a better control of the propagation of this measurement noise as well as of the usual measurement noise outside of the integral sign.

## 1. Deconvolutions and Image reconstruction

### 1.1. Maps of radio emission and the CLEAN algorithm

Probably the most well known inverse problem in astronomy is that of reconstructing the spatial structure of a source of radio wavelength radiation from observations with interferometric radio arrays, also known as radio synthesis mapping. There are a number of books on the technical aspects of radio wavelength observations and radio interferometry (e.g. Christiansen & Högbom, 1985 ; Thompson, Moran, & Swenson, 1986), so only a brief overview is given here.

There are various ways of setting up an array of radio telescopes such that they can be used as (a number of) interferometric pairs. Historically two configurations have been used. One is a linear array of telescopes that makes use of the rotation of the earth to obtain varying projections on the sky of the vectors connecting the individual telecopes. A single 'snapshot' interferogram of such a linear array has only one-dimensional spatial information. There is no sensitivity to structures along the direction perpendicular to the line along which the telescopes are arranged. The rotation of the earth makes the projection on the sky of this line rotate so that after 12 hours of gathering 'snapshots' all spatial directions are covered. The Westerbork Synthesis Radio Telescope (WSRT) in the Netherlands makes use of this technique. The advantage of such an array is that of all possible base lines between pairs of telescopes a large fraction re-occurs for different pairs.

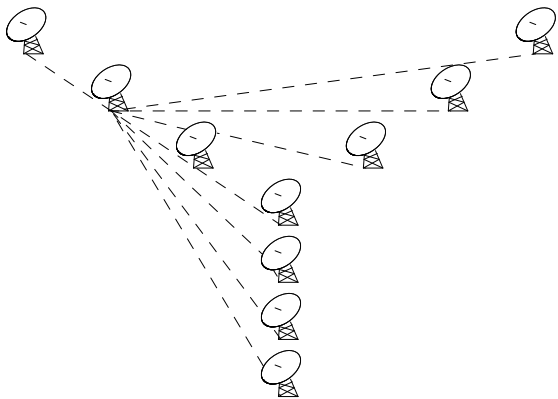

**Figure 1.** A sketch of a skeleton radio array. Dashed lines represent some of the base lines of this array. Since this array has three arms there is two-dimensional information in a single 'snapshot' so that it is not necessary to use the earth's rotation for synthesis of a radio image.

Such independent simultaneous measurements of the same spatial structure, commonly referred to as redundancy, can be used to improve signal to noise ratios and correct for various non-random instrumental effects such as due to uncertainties in the positioning of the telescopes. A detailed discussion of such an array and the data analysis procedures was presented by Brouw (1975).

The alternative is to build an array of telescopes in which the telescopes do not align, an example of which is shown in figure 1. Such an array has two-dimensional spatial information in a single 'snapshot' which means that the data gathering is much more efficient. There is much less redundancy for such a configuration than for a linear array, so this configuration places heavier demands on removing sources of non-random errors such as inaccurate measurements of the base lines between interferometer pairs. The Very Large Array (VLA) in New Mexico has a configuration similar to that shown in figure 1.

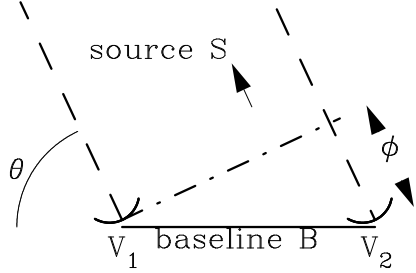

**Figure 2.** A simple two-element correlating interferometer. The base line is a vector of length $B$. The direction towards the source is the unit vector $s$. $\phi$ is difference in phase measured at the two elements due to the difference in path length, for an incoming plane wave.

To illustrate the techniques for building up images it is convenient to start by describing the image reconstruction for the simplest possible array, consisting of only two elements. This set-up is shown in figure 2. For a point source at infinity the incoming electro-magnetic waves are plane waves. The voltages measured at the radio antenna feeds are :

$$
\begin{aligned}
V_1 &\propto E\cos(\omega t) \\
V_2 &\propto E\cos(\omega t - \phi) \\
&= E\cos\left(\omega t - \frac{2\pi|B|}{\lambda}\cos\theta\right)
\end{aligned}
\quad (1)
$$

It is normal to record only the correlations i.e. the product of $V_1$ and $V_2$ using analog multiplication techniques. A high frequency term is filtered out immediately and the response $R$ then is :

$$R \propto S\cos\left(\frac{2\pi|B|}{\lambda}\cos\theta\right) \quad (2)$$

Here the flux density or power $S$ of the source has replaced the $E^2$ term. The $\cos\theta$ clearly comes from an inner product of the base line vector **B** and the vector towards the source **s**. The more general formulation in terms of this inner product of **B** (now measured in units of the wavelength of the radio waves) and **s** is then :

$$R \propto S\cos(2\pi\mathbf{B}\cdot\mathbf{s}) \quad (3)$$

It is clear that displacing the source along a direction perpendicular to **B** does not change the inner product $\mathbf{B}\cdot\mathbf{s}$ and therefore the response $R$ is invariant to such a displacement. This is the mathematical expression of having spatial resolution in only one direction for a linear array of radio telescopes.

For a source that is extended on the sky rather than a point source the simple product of $S$ and the phase delay term in equation (3) becomes a convolution integral :

$$R = \int d\sigma I(\sigma)\cos[2\pi\mathbf{B}\cdot(\mathbf{s}+\sigma)] \quad (4)$$

Here $I$ is the brightness distribution of the source modified by the response (beam pattern) $A(\sigma)$ of an individual radio antenna, so $I$ is itself a convolution of the intrinsic distribution and $A$. Now **s** is merely a convenient reference point near the source. Using the projected spacing **b** on the sky defined by

$$\mathbf{b} \equiv \mathbf{B} - (\mathbf{s} \cdot \mathbf{B})\mathbf{s} \qquad (5)$$

and also the approximation that the angular extent of the region observed at any given time is small so that $\sigma$ is nearly perpendicular to **s** the response reduces to :

$$R = V \exp[i2\pi \mathbf{B} \cdot \mathbf{s}] \qquad (6)$$

with the complex visibility $V$ defined by

$$V \equiv \int d\sigma I(\sigma) \exp[i2\pi \mathbf{b} \cdot \sigma] \qquad (7)$$

Since $\mathbf{B} \cdot \mathbf{s}$ is known, the visibility $V$ can be treated as the observable quantity. Typical radio telescope dishes are weighty metal structures of several tens of meters across and therefore their position in an array is generally fixed or at best they are movable only along a short rail track. This means that the number of available distinct base line vectors $\mathbf{B}_j$ or projected base lines $\mathbf{b}_j$ is small and fixed. This is one example of the impossibility in astronomy of controlling the sampling beyond the basic design of an array. The visibilities $V_j$ are measured for this finite (small) set of discrete projected base lines $\mathbf{b}_j$ in equation (7). The task at hand is then to reconstruct $I(\sigma)$ as closely as possible.

Equation (7) is a straightforward Fourier transform, which is discretized because of the discrete number of available base-lines. The inverse is then trivially :

$$I(x,y) = \sum_{j=1}^{N} w_j V(u_j, v_j) \exp[-i2\pi(u_j x + v_j y)] \qquad (8)$$

where $V(u_j, v_j)$ is the complex visibility function measured at base line $\mathbf{b}_j$ with $u_j$ and $v_j$ the two components of $\mathbf{b}_j$. The $(x,y)$ are the coordinates on the sky and the $w_j$ are weighting factors associated with the measured visibility function. Common choices are uniform weighting, which means that all visibilities are weighted equally, and natural weighting in which the $w_j$ are inversely proportional to the variance of the visibilities. In any case the weights must satisfy :

$$\sum_j w_j \equiv 1 \qquad (9)$$

The corresponding instrument response of the array (synthesized beam) is

$$P(x,y) = \sum_{j=1}^{N} w_j \exp[-i2\pi(u_j x + v_j y)] \qquad (10)$$

which should preferably be a localized function. In the field of optical image processing the true brightness distribution $I_0$ would be called the 'object'. $I$ would be called the 'image', but in radio astronomy it is generally referred to as 'dirty map' and the instrument response $P$ (point spread function) is referred to as 'dirty beam'. The term dirty is a graphic expression of the presence of unwanted secondary responses.

The two-dimensional discrete Fourier transforms (8) and (10) of the data are generally performed as a succession of one-dimensional fast Fourier transforms (FFTs). The problem with this direct approach lies in that often it is not possible with any set of weights $w_j$ to achieve a very nicely localized synthesized beam $P$. A regularly spaced array of telescope elements allows the use of a simple and efficient FFT but unfortunately also produces strong secondary response peaks of the array. A strong point source can therefore produce quite strong non-local structures in the image, usually referred to as grating rings or side-lobes, which then dominate the emission of potentially interesting faint extended features.

A very widely used technique to recognize side-lobes and eliminate their effect from the image is the iterative beam removing technique known as the CLEAN algorithm (Högbom, 1974). The CLEAN algorithm can be shown to be a least squares fit of harmonics in the Fourier domain (Schwarz, 1978). It is therefore not surprising that this method has also been applied in time series analysis (e.g. Roberts, Léhar, & Dreher, 1987). A more modern incarnation of the CLEAN algorithm is known as multi-resolution CLEAN or MRC (Brinks & Shane, 1984 ; Wakker & Schwarz, 1988), which itself recently has been modified by using wavelet transforms in combination with FFTs (cf. Starck et al., 1994). Both CLEAN and MRC are applied widely in radio synthesis mapping.

The basic CLEAN algorithm iterates over a number of reduction steps which are aimed essentially at representing the image as a collection of point sources with different intensities, convolved with the beam of the array. These successive steps are :

1. A search is conducted for the maximum in the correlation between the dirty map and the dirty beam. This is close to or even identical with the absolute largest value in the dirty map. The plausible assumption is made that this response is primarily due to a real point-source signal.
2. Some fraction $g$ of this peak value is accepted as the amplitude of the first delta function (component) from which the object is built up.
3. A dirty beam pattern scaled by this value and centered at the appropriate position is subtracted from the dirty map. This 'cleans up' the map because it removes a number of the unwanted secondary responses as well.
4. The remaining map after this subtraction is regarded as the new dirty map, and the iteration proceeds.

At some point the iteration process is stopped and a 'clean map' is usually constructed by convolving the components obtained with a hypothetical 'clean beam' (i.e. one without side-lobes) and adding the residuals from the final iteration step. The 'clean beam' can be e.g. a Gaussian with a width determined by the highest achievable spatial resolution, which corresponds to the longest base line in the array. Details of the criteria for ensuring convergence and criteria for halting the iteration process can be found in the papers in which the CLEAN algorithm is presented (Högbom, 1974) and developed (Schwarz, 1978 ; Wakker & Schwarz, 1988).

The main problem with the CLEAN algorithm, and the reason for developing the MRC algorithm, is that CLEAN is not suited to reconstructing very extended sources of radio emission. The reason for this is that spatial structure which extends over scales that are large compared with the resolution of the shortest base line in the array is not detected :

too little of the Fourier transform of this structure is sampled by the array. Expressing the same in other words : smooth extended structure can be only poorly approximated by a set of delta functions. The MRC solution for this problem is to convolve the 'dirty map' with a very broad smoothing function. This 'smooth dirty map' is subtracted from the 'dirty map' to obtain a 'difference dirty map'. The 'smooth dirty map' and the 'difference dirty map' together contain the same information that the original dirty map did. However the extended structure in the original map is much more 'point-source like' in the degraded resolution in the 'smoothed dirty map', so it can be represented much more easily with a delta function. Each of these two maps are now separately CLEANed according to the original algorithm and the results are added. The details of the algorithm and some tests can be found in the paper by Wakker & Schwarz (1988). This idea can of course be extended to more levels of successively degraded resolution which would be a true *multi*-resolution CLEAN rather than merely a dual resolution CLEAN. Some steps in this direction are taken in the paper by Starck et al. (1994) who also use wavelet transforms for the scaling between the resolution levels. The interested reader is referred to this paper for details.

*1.2. Statistical estimators : the MEM and Lucy's algorithm*

Consider again the typical linear inversion problem in astronomy leading to a Fredholm equation of the first kind :

$$g(t) = \int_a^b ds\ K(t,s) f(s), \qquad (11)$$

where $t$ and $s$ can be vectors. Equivalently one can consider the discrete (discretized) analogue :

$$\mathbf{g} = \mathbf{K} \cdot \mathbf{f} \qquad (12)$$

Since in astronomy the measurement errors are always large enough to be important in the inversion process it is perhaps inappropriate to treat the inversion in the classical fashion where an inverse exists and is unique as long as $\mathbf{K}$ is invertible. Even if the matrix $\mathbf{K}$ were not ill-conditioned the introduction of measurement errors in the data $\mathbf{f}$ will imply an uncertainty in the determination of $\mathbf{g}$. There is therefore never uniqueness in the mathematical sense. This is in itself not so important so long as reliable (finite) estimates of the uncertainty in the determination of $\mathbf{g}$ can be given. This is strictly a problem of statistical inference. Thus it is sensible to use a statistical method to attempt to find the most likely function $f(s)$ that satisfies the constraints posed by the data and possibly additional a-priori information. Two inversion schemes that take this maximum likelihood approach are in wide in use in astronomy. One is the Maximum Entropy Method (MEM), the other is known as Lucy's algorithm. A review of the use of the MEM in astronomy is the paper by Narayan & Nityananda (1986) and Lucy's algorithm is presented by Lucy (1974). The latter is developed in later papers by the same author (Lucy, 1992 ; Lucy, 1994), a variation on this idea can be found in a paper by Tsumuraya et al. (1994). The algorithms in use for the MEM have been developed primarily by Skilling, Gull and Bryan and descriptions can be found in a number of papers by these authors (cf. Skilling & Bryan, 1984 ; Skilling and Gull, 1985). Applications of the MEM can be found e.g. in papers by Horne (1985, 1994) and Marsh and Horne (1988).

The idea of the MEM is sufficiently well-known that only a very brief outline needs to be given here. The starting point is Bayes' relation between conditional probabilities :

$$P(A|B) = P(B|A)\frac{P(A)}{P(B)} \qquad (13)$$

Here $A$ and $B$ can be any statements or events to which a probability of occurrence can be assigned. In the current application $A$ is the unknown quantity or object, $B$ are the measured data (/image/visibility function). The conditional probability $P(A|B)$ is the probability of an object $A$ given the measurements $B$, which we need to maximize. $P(B|A)$ is the probability of finding measurements $B$ given object $A$. Adding the measurement noise term $N$ to the right-hand side of equation (12) and assuming it arises due to uncorrelated Gaussian random processes with variance $\sigma^2$ yields for $P(B|A)$ :

$$P(B|A) \propto \prod_j \exp\left(-\frac{N_j^2}{2\sigma_j^2}\right) \propto \prod_j \exp\left[-\frac{(\sum_i K_{ji}f_i - g_j)^2}{2\sigma_j^2}\right] \qquad (14)$$

The a-priori probability of the measurements $P(B)$ is independent of the unknown $A$ so in a maximization of the likelihood of $A$ it is unimportant except for normalization. The a priori distribution of objects $P(A)$ is where the entropy comes in.

$$P(A) \equiv P(f_i) \propto \exp S(f_i) \qquad (15)$$

Here $S(f_i)$ is the entropy of object $i$ which is a measure of the number of ways a given macroscopic state $\mathbf{f}$ can be built up from elementary events $\{i\}$. If these elementary events (atoms) $\{i\}$ each have a probability $p_i$ of occurring the entropy is :

$$S = -\sum_i p_i \ln p_i \qquad (16)$$

The advantage of defining the entropy like this is that it is additive for independent systems. The probabilities $p_i$ in the case of observing a number of photons which are distributed over a number of pixels/bins are the fraction in pixel $i$ of the total number of photons i.e. $p_i = f_i/\sum_i f_i$. Note that implicitly an extra assumption is made about the unknown function $f$ which is that it is positive semi-definite over its domain and that its integral is unity :

$$\int \mathrm{d}s\, f(s) = 1, \qquad f(s) \geq 0 \qquad (17)$$

In astronomy inversion problems for which such a-priori information is applicable or which can be slightly reformulated so that this applies (such as the deconvolution of images) are quite common so this is not considered to be a severe restriction and can even be an advantage.

Taking into account prior information such as symmetry properties of the images introduces degeneracy $q_i$ in the elementary states/events $\{i\}$ which leads to an entropy

$$S = -\sum_i p_i \ln(p_i/q_i) \qquad (18)$$

Combining (14) and (15) with (13) then leads to :

$$\ln P(A|B) = S(f_i) - \sum_j \left( \sum_i K_{ji} f_i - g_j \right)^2 / 2\sigma_j^2 \qquad (19)$$

which is the function to be minimized for the $f_i$. It is clear that equation (19) is a least squares or maximum likelihood estimator which is regularized by the non-linear entropy function $S$. Because this function is non-linear the minimization algorithm is not particularly simple. Estimating the propagation of measurement errors to the final result also is not a trivial matter. More discussion of this can be found in the papers by Narayan & Nityananda (1986) and Horne (1994).

Lucy's algorithm (1974) also uses equation (13) as its starting point to invert equation (11), and it also uses the a priori constraints (17). The notation used by Lucy is slightly different and in order to facilitate the discussion the following changes are made here :

$$\begin{cases} \mathbf{f} \text{ or } P(A) & \to \psi(\xi) \\ \mathbf{g} \text{ or } P(B) & \to \phi(x) \\ \mathbf{K} \text{ or } P(B|A) & \to P(x|\xi) \\ P(A|B) & \to Q(\xi|x) \end{cases} \qquad (20)$$

The new forms of (11) and (13) are

$$\begin{cases} \phi(x) = \int d\xi \, \psi(\xi) P(x|\xi) \\ Q(\xi|x) = \psi(\xi) P(x|\xi) / \phi(x) \end{cases} \qquad (21)$$

Re-arranging terms in the second of these equations and making use of the normalization of $P$ yields :

$$\psi(\xi) = \int dx \, \phi(x) Q(\xi|x) \qquad (22)$$

This is only apparently an analytical inverse with $Q$ as kernel because of course $Q$ depends on the unknown $\psi$ whereas the true inverse kernel would not. The reason to present this equation (22) is that it suggests an iterative procedure to determine $\psi$ as follows. Starting with an initial guess $\psi^0$, successive $\psi^r$ are calculated by evaluating in order :

$$\begin{aligned} \phi^r(x) &= \int d\xi \, \psi^r(\xi) P(x|\xi) \\ Q^r(\xi|x) &= \frac{\psi^r(\xi) P(x|\xi)}{\phi^r(x)} \\ \psi^{r+1}(\xi) &= \int dx \, \widetilde{\phi}(x) Q^r(\xi|x) \end{aligned} \qquad (23)$$

where $\widetilde{\phi}$ is the approximation to $\phi$ using the observational data. The second and third of these equations can be combined to eliminate $Q^r$ :

$$\psi^{r+1}(\xi) = \psi^r(\xi) \int dx \, \frac{\widetilde{\phi}(x)}{\phi^r(x)} P(x|\xi) \qquad (24)$$

Note that at each step the normalization and the positive semi-definite properties of $\psi$ are guaranteed.

In this iteration scheme corrections are applied on successively smaller spatial scales so that the scales which are most contaminated by noise are smoothed away. This iterative scheme always converges to the Maximum Likelihood solution although its rate of convergence can be quite slow. Accelerated schemes and stopping criteria for the iteration are discussed in the papers of Lucy (1992, 1994). Applications and variations of the scheme can be found in papers by e.g. Binney & de Vaucouleurs (1981), by Richichi et al. (1988), and by Tsumuraya et al. (1994).

## 2. MOLA and SOLA

### 2.1. the core of the algorithm

The method of subtractive optimally localized averages (SOLA) was developed with helioseismological applications in mind (cf. Pijpers & Thompson, 1992, 1994). The best way to compare SOLA with the older methods is probably by their application to the same problem, i.e. helioseismology. A review of the use of other inversion methods in this field, such as the non-linear invertible Abel transform which arises from asymptotic oscillation theory and the regularized least squares method, can be found in papers by Gough (1985), by Christensen-Dalsgaard et al. (1990), and by Schou et al. (1994).

The field of helioseismology concerns itself with deducing the internal structure of the Sun from the pulsation frequencies in particular of the oscillation modes known as the '5-minute oscillations'. Different modes of pulsation traverse different parts of the Sun and therefore sense the properties of the Sun in different ways. Deducing e.g. the run of the sound speed or the solar differential rotation can be reduced to a classical inversion of an integral equation. As an example and to introduce the method the inversion for the solar differential rotation is presented here.

Under the standard assumptions of linear stellar pulsation theory (cf. Cox, 1980) the wave equation that describes the amplitude of pulsation in stars belongs to the class of Sturm-Liouville boundary value problems. The amplitude of the pulsation of the Sun is small enough that it can be treated as linear. Therefore Sturm-Liouville theory applies and the solar pulsation can be decomposed uniquely into linearly superposed eigen-solutions or eigen-modes of the wave equation. The eigen-modes are usually separated in terms of functions depending on radius $r$ only and spherical harmonic functions $Y_l^m(\theta, \phi)$. Any mode can then be identified uniquely by its number of nodes $n$ in the radial direction and the order $l$ and degree $m$ of the spherical harmonic. The three components of the displacement vector $\xi$ around the mean state (hydrostatic equilibrium) are then :

$$\xi(r, \theta, \phi, t) = \left( \xi_r(r) Y_l^m, \ \xi_t(r) \frac{\partial Y_l^m}{\partial \theta}, \ \xi_t(r) \frac{1}{\sin\theta} \frac{\partial Y_l^m}{\partial \phi} \right) \\ \times \exp(-i\omega t) \qquad (25)$$

Here $\xi_r$ and $\xi_t$ are functions of radius $r$ only and are determined by solving a boundary value problem in the single independent variable $r$. With past and current observational facilities it is possible to detect in excess of 1000 distinct $(nl)$ multiplets and many or even all $m = -l, ..., l$ for these multiplets, which can thus amount to in excess of 60 000 modes in total.

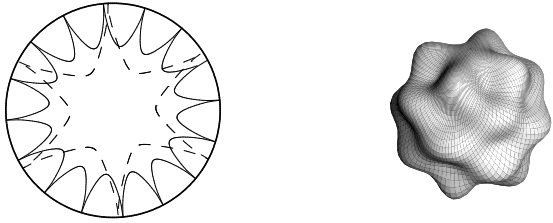

**Figure 3.** The left panel shows ray paths of seismic eigen-modes inside the body of the Sun for two different modes (dashed and full lines respectively). The right panel shows, with an exaggerated vertical scale, the surface deformation of the Sun for one eigen-mode : an $l = 8, m = 4$ spherical harmonic.

In a spherically symmetric, non-rotating star, the frequency $\omega_{nl}$ of a spheroidal mode of oscillation of radial order $n$ and degree $l$ is independent of the azimuthal degree $m$ of that mode. For this reason one can refer to the $(n, l)$ combination as multiplets. For a star that is rotating slowly the frequency is given by :

$$\omega_{nlm} = \omega_{nl} + m\Delta\omega_{nl} \qquad (26)$$

where $\Delta\omega_{nl}$ depends on the interior rotation rate $\Omega$ through an integral equation. This linear splitting of the frequencies is analogous to Zeeman splitting of multiplets in atomic physics. For the measured $\Delta\omega_{nl}$ the following holds :

$$\Delta\omega_{nl} = \int_{-1}^{1}\int_{0}^{1} dx\, d\cos\theta\; K_{nlm}(x,\theta)\Omega(x,\theta) + \epsilon_{nlm} \qquad (27)$$

Here $K_{nlm}$ is the rotational kernel for the mode, which is a known scalar function of the displacement vector $\xi$, and therefore can be treated as a known function of the fractional radius $x = r/R_\odot$ and co-latitude $\theta$. The $\epsilon_{nlm}$ are the measurement errors in the $\Delta\omega_{nl}$. Although full two-dimensional inversions are now carried out it is not unusual to invert for moments of the rotation rate with respect to $\cos\theta$ which reduces the two-dimensional problem to a one-dimensional inversion for the radial structure. For this reason but primarily for clarity the latitudinal dependence will be dropped here so that the SOLA inversion method is presented for a one-dimensional problem. The extension to more than one dimension is trivial although the notation gets more cumbersome.

In all OLA methods a set of coefficients is constructed such that the averaging kernel

$$\mathcal{K}(x_0, x) = \sum_{i \in \mathcal{M}} c_i(x_0) K_i(x) \qquad (28)$$

is peaked around $x = x_0$ and is small everywhere else. Here $\mathcal{M}$ is the set of observed oscillation modes and just a single subscript $i$ is used to represent the modes instead of the pair $(nl)$. A further constraint that is imposed on this averaging kernel is

$$\int_0^1 dx\; \mathcal{K}(x_0, x) = 1 \qquad (29)$$

An estimate of the rotation rate at $x = x_0$ can thus be obtained by defining

$$\langle \Omega(x_0) \rangle \equiv \sum_i c_i(x_0)\omega_{1i} = \int_0^1 dx \; \mathcal{K}(x_0, x)\Omega(x) + \sum_i c_i \epsilon_i \qquad (30)$$

The first term in this equation is a weighted average of $\Omega$ over the solar interior in which $\mathcal{K}$ is the weighting function. The other term is the error in this average propagated from the data errors.

In the classical OLA methods like the Backus & Gilbert method (1968) the coefficients $c$ are obtained by minimizing

$$\int_0^1 dx \; [\mathcal{K}(x_0, x)]^2 \mathcal{J}(x_0, x) + \mu \sum_{ij} E_{ij} c_i c_j \qquad (31)$$

subject to the constraint (29). $E$ is the error variance-covariance matrix of the data errors. $\mu$ is a free parameter that can be adjusted according to the relative desirability of limiting the magnification of the errors and localizing the integration kernel. The weighting function $\mathcal{J}$ is intended to ensure the localization of $\mathcal{K}$. A usual choice is the one of Backus & Gilbert (1968) : $\mathcal{J} = 12(x - x_0)^2$. The small value of $\mathcal{J}$ around $x = x_0$ ensures that a large value of $\mathcal{K}$ there will give a small contribution to the integral. A large value of $\mathcal{J}$ elsewhere produces a large contribution to the integral even for small values of the kernel $\mathcal{K}$. Therefore this optimization can produce a kernel with the required properties.

In this formalism the weighting function is multiplied by the function that must be localized in the OLA procedure so Pijpers & Thompson (1992, 1994) have started to refer to this as multiplicative OLA or MOLA. The alternative method proposed by them is to optimize

$$\int_0^1 dx \; [\mathcal{K}(x_0, x) - \mathcal{T}(x_0, x)]^2 + \mu \sum_{ij} E_{ij} c_i c_j \qquad (32)$$

Instead of multiplying by a given function $\mathcal{J}$ a target form $\mathcal{T}$ for the averaging kernel is subtracted. Thus if a localized $\mathcal{T}$ is introduced then departures of $\mathcal{K}$ from $\mathcal{T}$ are penalized in this optimization. One choice of $\mathcal{T}$ that works well in practice is

$$\mathcal{T} = \frac{1}{f\Delta} \exp\left[-\left(\frac{x - x_0}{\Delta}\right)^2\right] \qquad (33)$$

In this definition of a target function the constant factor $f$ is introduced to produce an integral of unity over the domain $[0, 1]$. For very small values of the width $\Delta$ this factor approaches $\sqrt{\pi}$. The width $\Delta$ is an adjustable parameter which specifies the required resolution. The two parameters $\mu$ and $\Delta$ must be adjusted to give acceptable matching of the averaging kernel to its target form and also an acceptably small error bound on the result from the propagated measurement errors. Note that there is no reason at all to use only this function (33) and it can be advantageous for some applications to use a completely different form.

The first advantage of SOLA methods over MOLA methods lies in the way that each equation is optimized. The MOLA method, i.e. minimization of equation (31), leads to a matrix inversion

$$\mathbf{A}^{MOLA}(x_0)c(x_0) = v \qquad (34)$$

where the vector $v$ is known and the matrix $\mathbf{A}$ has to be calculated and inverted for every $x_0$ because the individual elements of the matrix depend on $x_0$ through the function $\mathcal{J}$. In the SOLA method there is a similar matrix equation to be solved

$$\mathbf{A}^{SOLA} c(x_0) = v(x_0) \tag{35}$$

Now $\mathbf{A}$ is a matrix with the elements

$$\begin{aligned} A_{ij} &= \int_0^1 \mathrm{d}x \; K_i(x) K_j(x) + \mu E_{ij} \\ A_{i\,M+1} &= \int_0^1 \mathrm{d}x \; K_i(x) \\ A_{M+1\,j} &= \int_0^1 \mathrm{d}x \; K_j(x) \\ A_{M+1\,M+1} &= 0 \end{aligned} \tag{36}$$

Here $M$ is the number of measured oscillation frequencies i.e. the number of modes used in the inversion. The $(M+1)^{th}$ row and column arise because of the constraint (29) which is incorporated using a Lagrange multiplier. Since the $K_i$ are known and the $E_{ij}$ can be estimated before carrying out the inversion, the matrix $\mathbf{A}$ has to be calculated only once. For a constant weighting $\mu$ it also has to be inverted only once. This means that the SOLA methods are significantly faster than the MOLA methods. The target function occurs only in the elements of the vector $v$

$$\begin{aligned} v_i &= \int_0^1 \mathrm{d}x \; K_i(x) \mathcal{T}(x_0, x) \\ v_{M+1} &= 1 \end{aligned} \tag{37}$$

Only this vector has to be calculated for every $x_0$ and then multiplied through the already inverted matrix $\mathbf{A}$.

The second advantage of SOLA lies in its versatility. In helioseismology the attainable resolution is better nearer to the solar surface, because the intrinsic wave length of the pulsational waves is smaller near the surface than in the deep interior. This can be used to adjust the width $\Delta$ as a function of $x_0$. Since the target function occurs only in the vector $v$ this is computationally no less efficient than using a constant $\Delta$.

Figure 4 (from Pijpers & Thompson, 1994) shows the constructed averaging kernels for a mode set which contains 834 p-mode oscillations, with frequencies in the range $2\,\mathrm{mHz} - 4\,\mathrm{mHz}$ and with degrees $l$ between 1 and 200. The errors were assumed to be uniform and uncorrelated so that $\mathbf{E} = \sigma^2 \mathbf{I}$. The mode set and parameters are realistic for real inversions for solar rotation. Each panel is marked with values for $\Lambda$, which is the factor by which the errors are magnified after the inversion, and $\chi$ which is the integral of the square of the mismatch between kernel and target.

$$\chi \equiv \int_0^1 \mathrm{d}x \; [\mathcal{K}(x_0, x) - \mathcal{T}(x_0, x)]^2 \tag{38}$$

This $\chi$ can be used to give bounds on the effect of this mismatch on the uncertainty in the result (cf Pijpers & Thompson, 1994).

$$\left| \int \mathrm{d}x \; [\mathcal{K} - \mathcal{T}] \Omega \right| \leq \chi^{1/2} \frac{\Omega_{\max} - \Omega_{\min}}{2} \tag{39}$$

Here $\Omega_{\min}$ and $\Omega_{\max}$ are the minimum and maximum values respectively of $\Omega$ in the domain. Of course these values are usually not known a-priori so one has to use the estimates from the resolved inversion or use other a-priori knowledge to arrive at these values. If this is done the strict upper limit may lose its mathematical rigour although in astrophysical applications the estimate can still be sufficiently accurate to be useful.

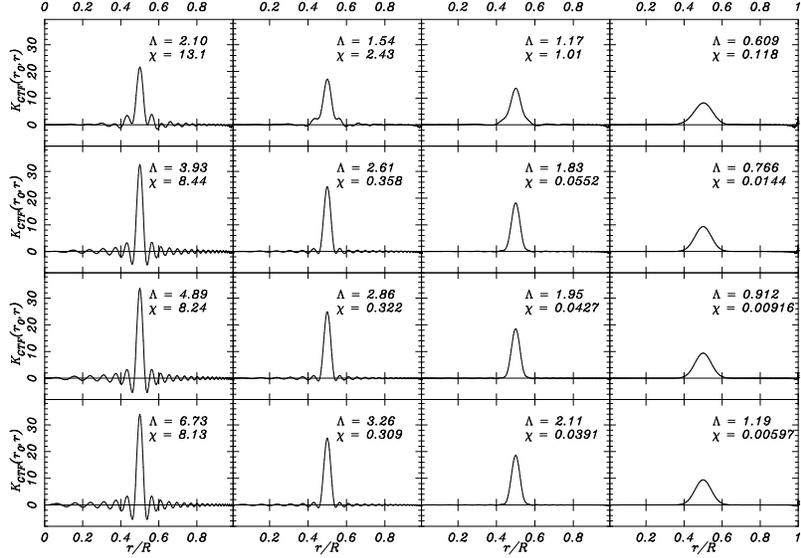

**Figure 4.** The constructed averaging kernels at radius $x = 0.5$ for various values of the width parameter $\Delta$ and the error weighting parameter $\mu$. $\Delta$ increases from left to right with values $10^{-2}$, $2\,10^{-2}$, $3\,10^{-2}$, $6\,10^{-2}$. The value of error weighting increases from bottom row up with values $3\,10^{-3}$, $10^{-2}$, $6\,10^{-2}$, $1.3$.

It is clear from figure 4 that if the widths $\Delta$ are chosen too small, an increased error weighting suppresses somewhat the side lobes of the kernel but not enough to give a satisfactory kernel. There are quite clear limits to the attainable resolution which are intrinsic to the mode set used.

Once the coefficients $c_i$ are obtained it the inner product of the coefficient vector $\mathbf{c}$ and the data vector $\Delta\omega$ yields an estimate of the rotation rate of the Sun $\langle\Omega(x_0)\rangle$ at the fractional radius $x_0$ :

$$
\begin{aligned}
\sum_i c_i \Delta\omega_i &= \int_0^1 \mathrm{d}x \, \sum_i K_i(x)\Omega(x) + \sum_i c_i \epsilon_i \\
&= \int_0^1 \mathrm{d}x \, \mathcal{K}(x_0,x)\Omega(x) + \sum_i c_i \epsilon_i \\
&\approx \int_0^1 \mathrm{d}x \, \mathcal{T}(x_0,x)\Omega(x) + \sum_i c_i \epsilon_i \\
&= \langle\Omega(x_0)\rangle + \sum_i c_i \epsilon_i
\end{aligned}
\quad (40)
$$

Note that it is also particularly simple in OLA methods to obtain a measure of the uncertainty of this estimate of the rotation rate due to propagation of measurement errors :

$$\sigma^2(\Omega) = \mathbf{c}^T \cdot \mathbf{E} \cdot \mathbf{c} \qquad (41)$$

where **E** is the measuremement error variance/co-variance matrix, **c** is the column vector of linear coefficients and $\mathbf{c}^T$ its transpose.

## 2.2. alternative target functions

The functional form of $\mathcal{T}$ also can be used to invert e.g. for spatial gradients in the rotation rate directly, rather than to infer such gradients from the reconstructed rotation rate profile (cf. Pijpers & Thompson, 1993a), or to invert for the integral of the so-called transfer function which is of interest in the reverberation mapping of the broad-line region of active galactic nuclei (Pijpers & Wanders, 1994).

To take the latter example first, the integral $F$ of the unknown function $f$ in the Fredholm equation (11) is :

$$F(t) = \int \mathrm{d}s \ f(s) \tag{42}$$

The constant of integration is irrelevant for this discussion and is set to 0 for convenience. This equation is clearly the same as equation (11) if one sets $K(t,s) \equiv 1$. Therefore if the target function $\mathcal{T}$ is chosen $\mathcal{T} \equiv 1$ in equation (32) one should obtain an optimal estimate of the integral, i.e. the 0th moment, of the unknown function $f$. In practice one finds that the higher resolution one requires in the variable $s$ the larger the magnification of data errors becomes. The resolution of the target function $\mathcal{T} = 1$ is very low and therefore the integral $F$ can generally be determined much more accurately than resolved values for $f$ (cf. Pijpers & Wanders, 1994).

In the same vein one can try to determine higher moments of $f$. For example the first moment of $f$ :

$$M^1(f) \equiv \int \mathrm{d}s \ sf(s) \qquad \longleftrightarrow \qquad \mathcal{T} = s \tag{43}$$

In other applications such as in helioseismology it may be of interest to find the value of gradients of the unknown $f$. In particular in stellar interiors large spatial gradients of the rotation rate or of the sound speed point to interesting physical transport effects. This means that in these applications we are interested in the finding a localized kernel $\mathcal{T}$ for :

$$\left\langle \frac{\partial \Omega}{\partial x}(x_0) \right\rangle = \int_0^1 \mathrm{d}x \ \mathcal{T}(x_0, x) \frac{\partial \Omega}{\partial x} \tag{44}$$

Using partial integration leads directly to :

$$\int_0^1 \mathrm{d}x \ \mathcal{T}(x_0, x) \frac{\partial \Omega}{\partial x} = [\mathcal{T}(x_0, 1)\Omega(1) - \mathcal{T}(x_0, 0)\Omega(0)] - \int_0^1 \mathrm{d}x \ \frac{\partial \mathcal{T}(x_0, x)}{\partial x} \Omega \tag{45}$$

The boundary terms can generally be assumed to be negligible if the width $\Delta$ of $\mathcal{T}$ is chosen small enough. This means that in order to detect gradients one can use the following form for a target function :

$$\mathcal{T}^{(d)}(x_0, x) = \frac{2(x - x_0)}{f\Delta^3} \exp\left[-\frac{(x - x_0)^2}{\Delta^2}\right] \tag{46}$$

This technique has been used in test situations (Pijpers & Thompson, 1993a, see also figure 5) but on real data the error magnification was found to be too large for this to be much use at this moment. A future set of modes with smaller measurements errors and a larger number of modes for use in the inversion may well make this target function useful.

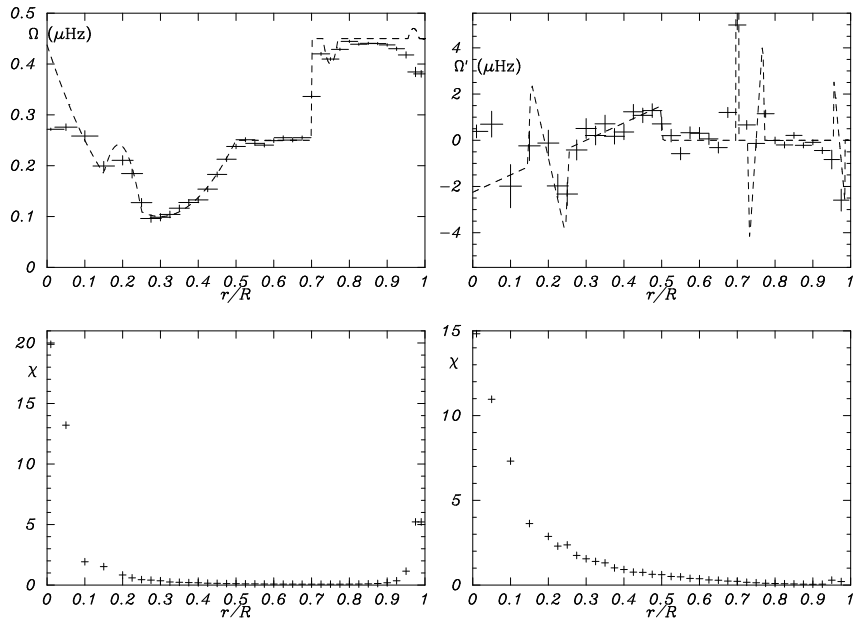

**Figure 5.** The upper left-hand panel shows the reconstruction of an artificial rotation rate as a function of radius inside the Sun using a set of 834 p-modes. The dashed line is the input, the crosses are the reconstructed values. The horizontal error bar shows the width of the kernel, the vertical error bar is the propagated data error. All quantities are realistic for inversions for solar rotation although the gradients in the profile of $\Omega$ are larger than expected. The lower left hand panel shows the associated values of $\chi$. Where this is large the reconstruction should not be trusted. The upper right hand panel shows the derivative of the rotation rate on the left and its reconstruction. The lower right hand panel shows the associated departure of constructed kernel from the target.

Naturally this idea can be extended to determinations for higher order derivatives. The practical use of this seems negligible because of the bad error propagation properties. However one regularization scheme in the regularized least squares methods minimizes the second spatial derivative of $\Omega$. In such a regularization scheme it is thus assumed that the second derivative of the reconstructed function is primarily due to noise. In the SOLA point of view one could equivalently determine the localized second derivative and subtract the appropriate contribution from the inversion in which (33) is used. In practice this procedure can be cut short by using the target form :

$$\mathcal{T}^{(reg)}(x_0, x) = \mathcal{T} - \frac{\Delta^2}{4}\frac{\partial^2 \mathcal{T}}{\partial x^2} \qquad (47)$$

with $\mathcal{T}$ as in equation (33). Using this as the target function the constructed averaging kernels $\mathcal{K}$ turn out to resemble closely the kernels from the least squares methods with second derivative smoothing regularization. They also have the property that the second moment around $x_0$ of these kernels is much smaller than for SOLA kernels using (33) as target function. It is simple to show that the target form (47) has a second moment around $x_0$ that is identical to 0. One should note however that higher order moments of $\mathcal{T}^{(reg)}$ are generally larger than higher order moments of $\mathcal{T}$. If one uses $\mathcal{T}^{(reg)}$ one must assume that the higher order moments of $\Omega$, whether real or introduced by errors, vanish sufficiently quickly to offset this property of the kernels $\mathcal{T}^{(reg)}$.

*2.3. Imperfectly known kernels : reverberation mapping*

The application of the SOLA method to the problem of reverberation mapping of the broad line region (BLR) of active galactic nuclei (AGN) is sufficiently different from the standard method described above that it is worth discussing separately. The BLR in AGN is too small to be resolved spatially with even the largest optical telescopes, so indirect ways must be found to obtain information about its structure and dynamics. The broad emission lines from which the BLR gets its name are photo-ionized by a small central continuum source. In many AGNs this continuum source is observed to be variable. The emission lines respond to this variation albeit with a time-delay of several days due to light-travel-time effects. The combination of flux and profile variations of the emission lines in response to the ionizing-continuum variations can thus be used to map the phase space of the BLR. This was shown in the paper by Blandford & McKee (1982) and a review of the field is a paper by Peterson (1993).

This mapping problem can be reduced to the following inverse problem :

$$L(v,t) = \int d\tau \, \Psi(v,\tau) C(t-\tau) \qquad (48)$$

here $L(v,t)$ is the observed emission-line light curve, $v$ the projected velocity with $v = 0$ the line centre. $C(t)$ is the observed continuum light curve, and $\Psi(v,\tau)$ the unknown transfer function. A number of different inversion methods have been used to invert this equation including the MEM (cf. Horne, 1994).

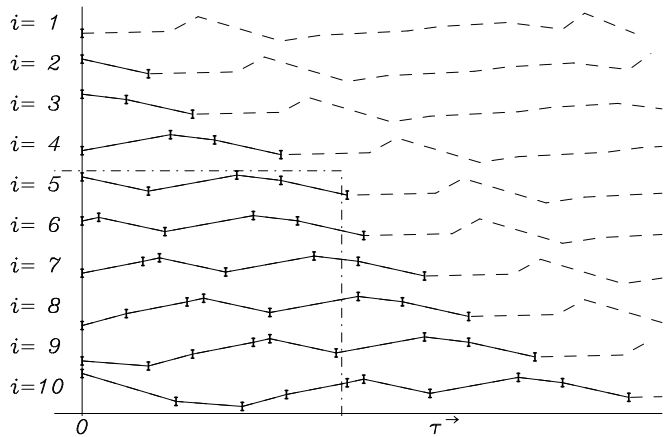

**Figure 6.** Example of a time series with 10 measurements. The index $i$ counts the consecutive measurements. The whole time series is replotted as a function of delay time $\tau$ for each added measurement, with an arbitrary vertical offset. The dashed lines represent the flux before the first measurement in the series.

The new aspect of this problem is that now the known function under the integral sign is not known as a continuous function with arbitrary precision over the entire domain of integration. Since the continuum is an observed quantity it is :

1. Sampled discretely at irregular intervals.
2. Sampled for a finite time span.
3. known with a finite precision, determined by instrumental effects.

The way to deal with these various problems within the SOLA framework is discussed in detail in the paper by Pijpers & Wanders (1994). Figure 6 (from Pijpers & Wanders, 1994) shows an example of a measured time series. For each measurement of the line-flux $L(t_i)$ there is a section of the measured continuum-flux time-series that lies in its past. It is usually assumed that the causality is such that the line-flux variations respond to the variations in the continuum and that therefore the lower limit of the integration in equation (48) can be set to 0. Furthermore it can be assumed that there is a maximum size to the BLR which, if no other arguments give a smaller upper limit, can always be set by the fact that it is not spatially resolved. This maximum size immediately gives a maximum to the time delay $\tau$ by dividing the maximum size by the velocity of communication : the speed of light. If the total length of the measured time series is smaller than this maximum time delay this means effectively that one cannot reasonably expect to do the inversion problem without doing something as arbitrary as extrapolating the time series. Since the pattern of variability does not show any single characteristic time scale such extrapolation is not constrained in any way. Therefore only the case for which the length of the time series is considerably longer than the maximum expected time delay is considered here.

In figure 6 the position of the maximum time delay $\tau_{\max}$ is denoted by the vertical dash-dotted line. All those measurements of the line flux $L(t_i)$ which have an associated partial continuum-flux time series with a length shorter than $\tau_{\max}$ would still require extrapolation to be usable for the inversion. Therefore these measurements are excluded, which is denoted by the horizontal dash-dotted line. It is clear from figure 6 that the remainder of the measurements of the emission-line flux, below and to the left of the dash-dotted lines, have associated continuum-flux time series that require only interpolation over the entire domain $[0, \tau_{\max}]$.

An interpolation scheme for the continuum is necessary within the SOLA framework because of the matrix elements that need to be calculated in equation (36). The partial time series of the continuum flux here play the role of the mode kernels in helioseismology so the interpolation is necessary to calculate all the cross products of these partial time series. Note that it is generally impossible to obtain a regularly spaced time series : nights of observing can be lost because of weather conditions, and seasonal gaps in the time series can arise if the object is only in the visible night sky for parts of the year. This means that calculating the required integrals in equation (36) using Fourier transform techniques is not a well defined procedure. Instead an interpolation scheme is used that is based on the Savitzky-Golay technique of fitting a low-order polynomial to a moving window of points in the time series (cf. Press et al., 1992).

The size of the window and the order of the polynomial used should depend on the ratio of the real variations in the continuum and the measurement noise. So far the procedure that has been followed is a relatively naive ad-hoc procedure described in the paper by Pijpers (1994).

The Savitzky-Golay fitting procedure is effectively a local filter suppressing high-frequency signal and therefore it affects the time signature of the continuum. This means it also has an influence on the correlation between line and continuum flux and the reconstructed transfer function. In practice one cannot expect to get an accurately resolved reconstruction if the resolution width $\Delta$ in the target function (33) is chosen smaller than the width of the window in the Savitzky-Golay interpolation scheme.

With these ingredients the SOLA method can proceed as discussed before. Thus a transfer function is obtained but the calculation of the uncertainty due to measurement errors is still to be carried out. The treatment of the measurement errors of the emission-line flux is the same as before. The calculation of an error in the transfer function from the propagated errors of the continuum-flux time-series is a complicated procedure because a matrix inversion is involved. In practice it is much simpler to allow the constructed integration kernel from the SOLA method to have an associated measurement error. This is simple to calculate since it is constructed from a linear combination of the interpolated partial continuum time-series. From this it is possible to give an error estimate for the transfer function $\Psi$ because it is possible to use the departure from the target form of the constructed averaging kernel (cf. Pijpers & Wanders, 1994).

A special case that is of particular interest occurs if the emission-line flux and the continuum flux are related by a simple phase delay :

$$L(t_i) = \text{const} \times C(t_i - t_d) \tag{49}$$

It is easy to see that this can be written in the same form as equation (48) where the transfer function $\Psi$ must then be :

$$\Psi = \text{const} \times \delta(\tau - t_d) \tag{50}$$

In this case the transfer function is completely determined by its zero-order moment $M^0$, which gives the value of the multiplicative constant, and its first moment $M^1$ :

$$M^1(\Psi) = M^0(\Psi) \times t_d \tag{51}$$

Now one should recall that these two low-order moments can generally be recovered with a higher accuracy than can the resolved transfer function $\Psi$ by using the inversions with $\mathcal{T} \equiv 1$ and the target function of equation (43). This means that even though the transfer function is much narrower than the resolution for the averaging kernels that can reasonably be expected, its position can be determined quite well by using a type of kernel that is specifically designed for this problem. It is not always necessary or even desirable to try to obtain a resolved result for the unknown quantity. A different method for the same problem, that is based on statistical estimation can be found in a paper by Rybicki & Press (1992).

## 3. Summing up

In astronomy there are a number of different problems in which the inversion of integral equations plays a rôle. From the brief overview given here one can see that the methods that have been designed to deal with each problem often make use of a-priori knowledge or assumptions to regularize the method and stabilize it against propagation of measurement errors and sampling deficiencies. As a consequence these methods are not always generally applicable to any inversion problem because what is considered their strength in the problem they are designed for could be a weakness in another.

For the same reason especially the non-linear methods are quite difficult to compare from an algorithmic point of view without actually discussing their application to one and the same inversion problem. For linear methods at least it is possible to compare their performance by examing the averaging kernels and the error propagation using the linear coefficients that the inversion algorithms produce. Thus it is possible, as shown in section 2.2, to translate the least squares method with second derivative smoothing regularization into the SOLA framework using a specific kernel. At least in principle the same should be possible for any linear regularization scheme.

The SOLA method shares the advantages of all linear methods in that its use is transparent and the treatment of measurement errors is straightforward. It has the same advantage of the classical Backus & Gilbert method in its use of the trade-off between resolution and error magnification without assuming anything about the unknown quantity under the integral sign. However it is more versatile in the choices of target function and distinctly more efficient in computation than the original Backus & Gilbert method.

**Acknowledgments :** I would like to thank the organizers of the Institute of Mathematics and its Applications workshop on Inverse Problems in Wave Propagation and in particular prof. Paul Sacks, for their invitation to present my work, and their generous hospitality during my stay.